\documentclass{article}

    \PassOptionsToPackage{numbers, compress}{natbib}

     \usepackage[preprint]{neurips_2023}

\usepackage[utf8]{inputenc} 
\usepackage[T1]{fontenc}    
\usepackage{hyperref}       
\usepackage{url}            
\usepackage{booktabs}       
\usepackage{amsfonts}       
\usepackage{nicefrac}       
\usepackage{microtype}      
\usepackage{xcolor}         
\usepackage{amsmath, bm}
\usepackage{graphicx}
\usepackage{wrapfig}

\usepackage{amssymb}
\usepackage{mathtools}
\usepackage{amsthm}
\usepackage{multirow}
\usepackage{subfigure} 
\usepackage{float}
\usepackage{algorithm} 
\usepackage{algorithmic}
\usepackage{textcomp} 
\usepackage{listings} 
\usepackage{enumitem}
\usepackage{xcolor}
\usepackage{natbib}

\title{Make-A-Voice: Unified Voice Synthesis With Discrete Representation}

\author{%
  Rongjie Huang$^{1}$\thanks{~~Equal contributions}, Chunlei Zhang$^{2}$\footnotemark[1], Yongqi Wang$^{1}$\footnotemark[1], Dongchao Yang$^{3}$, \\ \textbf{Luping Liu$^{1}$, Zhenhui Ye$^{1}$, Ziyue Jiang$^{1}$, Chao Weng$^{2}$, Zhou Zhao$^{1}$, Dong Yu$^{2}$} \\  \\ 
  Zhejiang University$^1$, Tencent AI Lab$^2$, Peking University$^3$ 
  }

\begin{document}

\maketitle

\begin{abstract}
Various applications of voice synthesis have been developed independently despite the fact that they generate ``voice'' as output in common. In addition, the majority of voice synthesis models currently rely on annotated audio data, but it is crucial to scale them to self-supervised datasets in order to effectively capture the wide range of acoustic variations present in human voice, including speaker identity, emotion, and prosody. In this work, we propose Make-A-Voice, a unified framework for synthesizing and manipulating voice signals from discrete representations. Make-A-Voice leverages a ``coarse-to-fine'' approach to model the human voice, which involves three stages: 1) semantic stage: model high-level transformation between linguistic content and self-supervised semantic tokens, 2) acoustic stage: introduce varying control signals as acoustic conditions for semantic-to-acoustic modeling, and 3) generation stage: synthesize high-fidelity waveforms from acoustic tokens. Make-A-Voice offers notable benefits as a unified voice synthesis framework: 1) Data scalability: the major backbone (i.e., acoustic and generation stage) does not require any annotations, and thus the training data could be scaled up. 2) Controllability and conditioning flexibility: we investigate different conditioning mechanisms and effectively handle three voice synthesis applications, including text-to-speech (TTS), voice conversion (VC), and singing voice synthesis (SVS) by re-synthesizing the discrete voice representations with prompt guidance. Experimental results demonstrate that Make-A-Voice exhibits superior audio quality and style similarity compared with competitive baseline models. \footnote{Audio samples are available at \url{https://Make-A-Voice.github.io}}

\end{abstract}

\section{Introduction}
Voice synthesis~\citep{wang2017tacotron,ren2019fastspeech,qian2020unsupervised} aims to generate human-like voices, which attracts broad interest in the machine learning community. A growing number of applications, such as voice assistant services and long-form reading, have been actively developed and deployed to real-world speech platforms. Despite the success achieved, the rising demand for expressive voice generation poses challenges for models, particularly in zero-shot scenarios~\citep{casanova2022yourtts,huang2022generspeech}. When the distributions of zero-shot style prompts differ from training data, the quality of synthesized voice often deteriorates due to distribution gaps. 

A promising approach to improve zero-shot robustness is to scale up training data with a large number of speakers, various accents, diverse demographics, and heterogeneous recording conditions to capture the acoustic diversities (e.g., intrinsic properties like speaker identity, emotion and prosody, and extrinsic factors such as channel and environment) in human voice. Current large-scale voice synthesis~\citep{kharitonov2023speak,wang2023neural,shen2023naturalspeech} systems leverage the codec models to predict the intermediate representation instead of waveforms, which can be clustered into two groups: 1) autoregressive (AR)~\citep{borsos2022audiolm,kharitonov2023speak,wang2023neural,zhang2023speak}: which integrates implicit duration modeling but suffers from error propagation. 2) non-autoregressive (NAR)~\citep{yang2023instructtts,shen2023naturalspeech}: which shows advantages in training stability but has been concerned with the requirement of an alignment model and ``averaged'' duration being predicted. Given that previous AR models~\citep{wang2023neural,zhang2023speak} have addressed intelligibility challenges in generated samples through large-scale training, we have incorporated the autoregressive approach into our model design.

While previous voice synthesis models have improved zero-shot robustness, most of them have been developed independently despite generating ``voice'' as a common objective. Thus, the methodologies developed for each application remain scattered in research fields, which is inefficient since we still need to optimize separated models for voice generation tasks.

In this work, we introduce Make-A-Voice, a unified zero-shot voice synthesis framework for synthesizing and manipulating voice signals from discrete representations. Make-A-Voice employs a ``coarse-to-fine'' design to model human voice, utilizing two types of discrete tokens: semantic tokens and acoustic tokens. This design encompasses the following stages: 1) Semantic stage is regarded as a sequence-to-sequence (seq2seq) problem which requires a small amount of text-audio parallel data; 2) Acoustic stage is scaled to a large amount of self-supervised audio-only data, where conditioning mechanisms are investigated with different control signals (e.g., speaker/F0 prompt); and 3) Generation stage is designed to reconstruct high-fidelity waveforms with accurate content (high intelligibility), rich acoustic properties.

Make-A-Voice demonstrates notable advantages as a unified voice synthesis framework: 1) Data scalability: the major backbone of Make-A-Voice (i.e., both acoustic and generation stages) does not require any annotations, and thus training could be scaled up in terms of data usage. 2) Controllability with flexible conditioning options: various conditioning mechanisms are investigated by re-synthesizing the semantic or acoustic representations with prompt guidance (e.g., speaker/F0). After training the backbone network, we proceed to introduce three voice generation applications: text-to-speech (TTS), voice conversion (VC), and singing voice synthesis (SVS). These applications can be effectively addressed by leveraging a unified framework that employs discrete representations. Experimental results demonstrate that Make-A-Voice achieves new state-of-the-art results in zero-shot synthesis. Both subjective and objective evaluation metrics show that Make-A-Voice exhibits superior audio quality and style similarity compared with baseline models. 

Our contributions can be summarized as follows:
\begin{itemize}[leftmargin=*]
    \item We propose a unified voice synthesis framework called Make-A-Voice, which stands for a three-stage approach with ``coarse-to-fine'' design. This framework aims to effectively model the human voice by considering semantic meanings, acoustic conditions, and perceptual fidelity.
    \item We enhance the scalability and leverage self-supervised data for both speech and singing voice synthesis.
    \item Experimental results on three exemplar applications demonstrate that Make-A-Voice achieves state-of-the-art results in terms of style similarity and perceptual quality. Make-A-Voice excels in data scalability, controllability, and conditioning flexibility with discrete representations.
    
\end{itemize}
\section{Related Works}
\subsection{Speech/Singing Synthesis}

TTS or SVS models typically convert input text into mel-spectrogram (e.g., Tacotron~\cite{wang2017tacotron}, FastSpeech~\cite{ren2019fastspeech}), which is then transformed to waveform using a separately trained vocoder~\cite{kong2020hifi,huang2021multi}, or directly generate waveform from text (e.g., EATS~\cite{donahue2020end} and VITS~\cite{kim2021conditional}). In zero-shot scenarios, when the distributions of style prompts deviate from the training data, the quality of the synthesized voice often suffers degradation due to distribution mismatches: Meta-StyleSpeech~\cite{min2021meta} generally adopts a speech encoding network for multi-speaker synthesis. GenerSpeech~\citep{huang2022generspeech} leverages multi-level style adaptors for the global and local stylization of the custom utterance. YourTTS~\citep{casanova2022yourtts} is built upon VITS with several novel modifications for zero-shot multi-speaker and multilingual training. In this work, we enhance zero-shot robustness by scaling up training data with an extensive collection of speakers encompassing diverse accents, demographics, and recording conditions. This approach aims to capture the acoustic diversity presented in human speech, including variations of speaker identity, emotion, and prosody. 

Building a simple and unified voice synthesis framework has also attracted increasing attention in the community: ~\citet{liu2023unifyspeech} make use of large amounts of unlabeled data for model training and boost the performance of zero-shot text-to-speech and voice conversion simultaneously. NANSY families~\citep{choi2021neural,choi2022nansy++} are trained in a self-supervised manner that does not require any annotations paired with audio. However, these methods are built in continuous vector space, easily encountering over-smoothing predictions. In contrast, our investigation focuses on the progressive development of a ``coarse-to-fine'' approach, which enables the synthesis and manipulation of voice signals based on discrete representations. 

\subsection{Self-Supervised Learning in Speech}

Self-supervised learning (SSL) has emerged as a popular solution to many speech processing problems with a massive amount of unlabeled speech data. Data2vec~\citep{baevski2022data2vec} employs a fast convolutional decoder and models the contextualized target representations in a self-supervised manner. HuBERT~\citep{hsu2021hubert} is trained using a masked prediction objective that involves masking segments of continuous audio signals. Additionally, it incorporates an offline clustering process to obtain aligned target labels for a prediction loss similar to BERT~\cite{devlin2018bert}. SoundStream~\citep{zeghidour2021soundstream} and Encodec~\citep{defossez2022high} draw inspiration from vector quantization (VQ) and investigate to use of a hierarchical architecture for representing acoustic information. \citep{yang2023hifi} propose a group-residual vector quantization (GRVQ) technique and use it to develop a high-fidelity audio codec model named HiFi-Codec. In this work, we leverage the semantic tokens from HuBERT and acoustic tokens from SoundStream as discrete representations.

\subsection{Spoken Language Models}

In a compact and discrete space, voice could be modeled with an autoregressive transformer. Generative spoken language modeling (GSLM)~\citep{kharitonov2022textless} with ``textless NLP'' pioneers the use of language models on discrete speech representations. AudioLM~\citep{borsos2022audiolm} and MusicLM~\citep{agostinelli2023musiclm} cast audio synthesis as a language modeling task and leverage a hierarchy of coarse-to-fine units. SpeechDLM~\citep{nguyen2023generative} leverages the HuBERT representation and introduces the end-to-end spoken language model for dialogue. Recently, VALL-E~\citep{wang2023neural}, SPEAR-TTS~\citep{kharitonov2023speak} are proposed to clone a human's voice with discrete prompt tokens from a short recording (3-seconds). In this study, our unified voice synthesis framework defines the voice generation process into a composition of autoregressive sequence-to-sequence (seq2seq) tasks utilizing discrete representations: translating speech/text to semantic tokens; converting semantic tokens to acoustic tokens; finally, mapping them back to long continuous waveforms. 


\section{Make-A-Voice}

\subsection{Zero-Shot Formulation}
Following~\citep{casanova2022yourtts,huang2022generspeech,wang2023neural}, we aim to generate high-fidelity samples with style (e.g., speaker identity, emotion, and prosody) guided by a prompt in a \textbf{zero-shot} manner, where model never sees the voice used for prompting at training, and it has to reproduce the acoustic characteristics from a single prompt example.

\subsection{Overview}

Make-A-Voice is considered a unified voice synthesis framework with a ``coarse-to-fine'' design that progressively enhances the modeling of voice signals by injecting desired conditioning information, which is organized in three main stages as illustrated in Figure~\ref{fig:arch}: 1) semantic stage $S_1$, speech or text inputs are transformed into a sequence of semantic tokens $s$, 2) acoustic stage $S_2$, acoustic tokens $a$ with a variety of conditions (speaker, emotion, prosody, and style) are generated autoregressively from the ``pseudo'' text (i.e., semantic tokens $s$); 3) generation stage $S_3$, a unit-based vocoder synthesizes high-fidelity waveforms from compressed acoustic representations. 


After training the backbone network, we introduce three voice generation applications, including text-to-speech (TTS), voice conversion (VC), and singing voice synthesis (SVS), that can be tackled by sharing a voice synthesis framework with discrete representations. For TTS, the text inputs are transformed into semantic tokens autoregressively, which are then converted into acoustic tokens given the speaker prompts. For VC, the speech inputs are tokenized into semantic tokens by the HuBERT model, which are then converted into acoustic tokens given the speaker prompts. For SVS, the semantic tokens are generated autoregressively as TTS, after which acoustic tokens are generated given the speaker and F0 prompts.


\begin{figure*}[]
    \centering
    \includegraphics[width=1.0\textwidth]{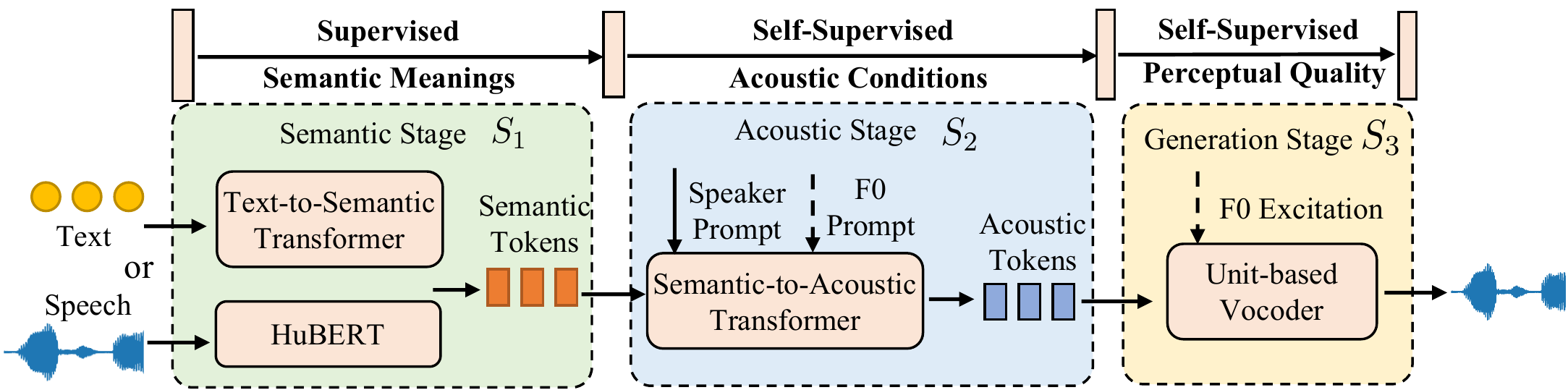}
    \caption{A high-level overview of Make-A-Voice. The F0 auxiliary input denoted with dotted lines is included only in singing voice synthesis.}
    \label{fig:arch}
  \end{figure*}

\subsection{Discrete Speech Representation}

\subsubsection{Semantic tokens}
\begin{wrapfigure}{r}{4cm}
    \centering
    \vspace{-8mm}
    \includegraphics[width=0.3\textwidth]{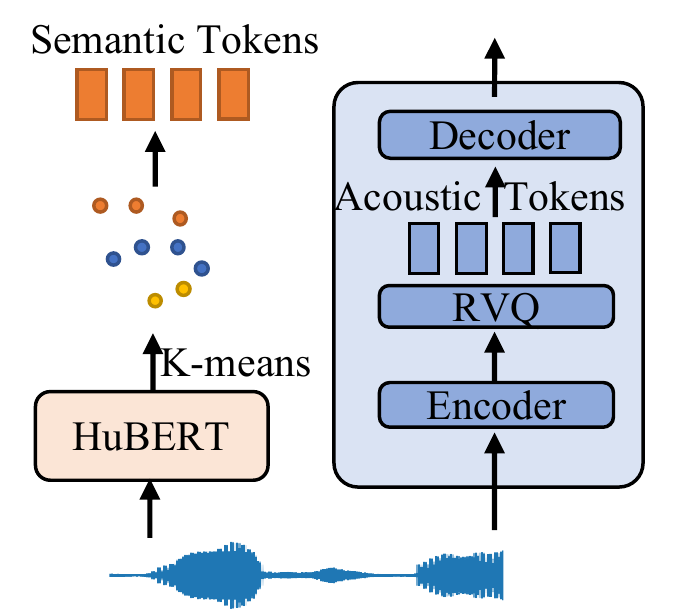}
    \vspace{-6mm}
    \caption{Discrete speech representations.}
     \vspace{-4mm}
    \label{fig:discrete}
\end{wrapfigure}


Discrete HuBERT~\citep{hsu2021hubert} units obtained from the clustering of self-supervised speech representations are shown~\citep{lee2021textless,lian2022utts} to be effective in providing linguistic content information. HuBERT trained on an unlabelled speech corpus encodes the target speech (16 kHz) into continuous representations at every 20-ms frame. A k-means algorithm is applied to the learned representations of the unlabelled speech to generate $K_1$ cluster centroids. In the end, a speech utterance $y$ is represented as semantic tokens with $\left[s_1, s_2, \ldots, s_T\right], s_i \in\{0,1, \ldots, K_1-1\}, \forall 1 \leq i \leq T$ , where $T$ is the number of frames.

\subsubsection{Acoustic tokens}
Audio codec models such as SoundStream~\citep{zeghidour2021soundstream} and Encodec~\citep{defossez2022high} have recently shown that encoder-decoder architecture excels at learning acoustic information in a self-supervised manner, where the representation can be used in a variety of generative tasks.

The acoustic codec model typically consists of an audio encoder, a residual vector-quantizer (RVQ), and an audio decoder: 1) The audio encoder $E$ consists of several convolutional blocks with a total downsampling rate of 320 and generates continuous representations at every 20-ms frame in 16kHz. 2) The residual vector-quantizer $Q$ produces discrete representations $a_q$ with a codebook size of $K_2$, using a vector quantization layer~\citep{vasuki2006review}. 3) The audio decoder $G$ reconstructs the signal $\hat{y}$, from the compressed latent representation $a_q$. In the end, a speech utterance $y$ is represented as acoustic tokens with $\left[a_1, a_2, \ldots, a_T\right], a_i \in\{0,1, \ldots, K_2-1\}, \forall 1 \leq i \leq T$ , where $T$ is the number of frames.

\subsection{Semantic Stage $S_1$: Determining Semantic Meaning}
Semantic stage $S_1$ maps tokenized text or speech into semantic tokens. For text-to-semantic transformation, we use parallel text-semantic data to learn this mapping with a sequence-to-sequence (seq2seq) task, that can be implemented by autoregressive encoder-decoder transformer architectures $\theta_{A R}$. It is conditioned on the phoneme sequence $u$, formulated as:
\begin{equation}
    p\left(\mathbf{s} \mid \mathbf{u}; \theta_{A R}\right)=\prod_{t=0}^T p\left(\mathbf{s}_{t} \mid \mathbf{s}_{<t}, \mathbf{u} ; \theta_{A R}\right),
\end{equation}
For speech-to-semantic transformation, we tokenize speech using the HuBERT model without requiring seq2seq translation.

\subsection{Acoustic Stage $S_2$: Introducing Acoustic Conditions}
Acoustic stage $S_2$ maps semantics into acoustic tokens. To train this stage, we extract pairs of sequences of semantic and acoustic tokens from each utterance. $S_2$ is scaled up to a large amount of self-supervised audio-only data containing many speakers with various accents, diverse demographics, and heterogeneous recording conditions to improve the robustness in zero-shot scenarios. It is designed to include a variety of acoustic conditions (e.g., speaker, emotion, prosody, and style) on top of semantic meanings, and thus we investigate different conditioning mechanisms for controllability and flexibility.

To control the characteristics of the speaker's voice, a prompt for timbre guidance is required. During training, we randomly select two non-overlapping windows of speech from each example, and consider one of the windows as the prompt and the other as the target output. For the conditioning mechanism, we concatenate the sequences of acoustic tokens $a_p$ from prompt and semantic tokens $s$ from the target, between which we have a separating token:
\begin{equation}
    p\left(\mathbf{a} \mid \mathbf{s,a_p}; \theta_{A R}\right)=\prod_{t=0}^T p\left(\mathbf{a}_{t} \mid \mathbf{a}_{<t}, \mathbf{s,a_p} ; \theta_{A R}\right)
\end{equation}
For singing voice synthesis which requires accurate pitch control due to its strong condition nature, the fundamental frequency (F0) prompt is further required. In practice, F0 could be determined by a separately-trained neural network given MIDI score, and thus we directly take the F0 value as condition signals in acoustic model $S_2$ for simplification following~\citep{liu2022diffsinger}. Specifically, we 1) extract the $\mathbf{F_0}=\left(f_1, \ldots, f_{L}\right)$ using the YAAPT algorithm~\citep{kasi2002yet} from target singing voice with 320 hop size, and 2) quantize the F0 of each frame to 256 possible values in log-scale, where each element in $f_i$ is an integer $f_i \in\left\{1, . ., K_f\right\}, K_f=256$. To conclude, we concatenate the sequences of acoustic tokens $a_p$ from prompt, F0 prompt $\mathbf{F}$, and semantic tokens $s$ from the target, between which we have separating tokens. 
\begin{equation}
    p\left(\mathbf{a} \mid \mathbf{s,a_p,F}; \theta_{A R}\right)=\prod_{t=0}^T p\left(\mathbf{a}_{t} \mid \mathbf{a}_{<t}, \mathbf{s,a_p,F} ; \theta_{A R}\right),
\end{equation}
\subsection{Generation Stage $S_3$: Reconstructing High-Fidelity Waveforms}

Generation stage $S_3$ guarantees the fidelity of synthesized waveforms. Since the acoustic codec (i.e., SoundStream) leverages multiple (12) quantization levels to improve reconstruction quality, and thus a distinct drop in perceptual quality could be witnessed when reducing the codebook number during inference. We refer the reader to Section~\ref{ablation:vocoder} for a summary of our findings. 

To avoid quality degradation, we train a unit-based neural vocoder from scratch for the acoustic unit to waveform generation, which only requires three quantization levels to reconstruct high-fidelity waveforms. Inspired by BigVGAN~\citep{lee2022bigvgan}, the synthesizer includes the generator and multi-resolution discriminator (MRD). The generator is built from a set of look-up tables (LUT) that embed the discrete representation and a series of blocks composed of transposed convolution and a residual block with dilated layers. The transposed convolutions upsample the encoded representation to match the input sample rate, while the dilated layers increase the receptive field. More details have been included in Appendix~\ref{vocoder}

For speech generation, we train the vocoder with only the discrete unit sequences as input. For singing voice generation, we further include F0-driven source excitation to stabilize long-continuous waveforms generation following~\citep{liu2022diffsinger,huang2022singgan}.

\subsection{Training and Inference Procedures}

\begin{figure*}[]
    \centering
    \includegraphics[width=1.0\textwidth]{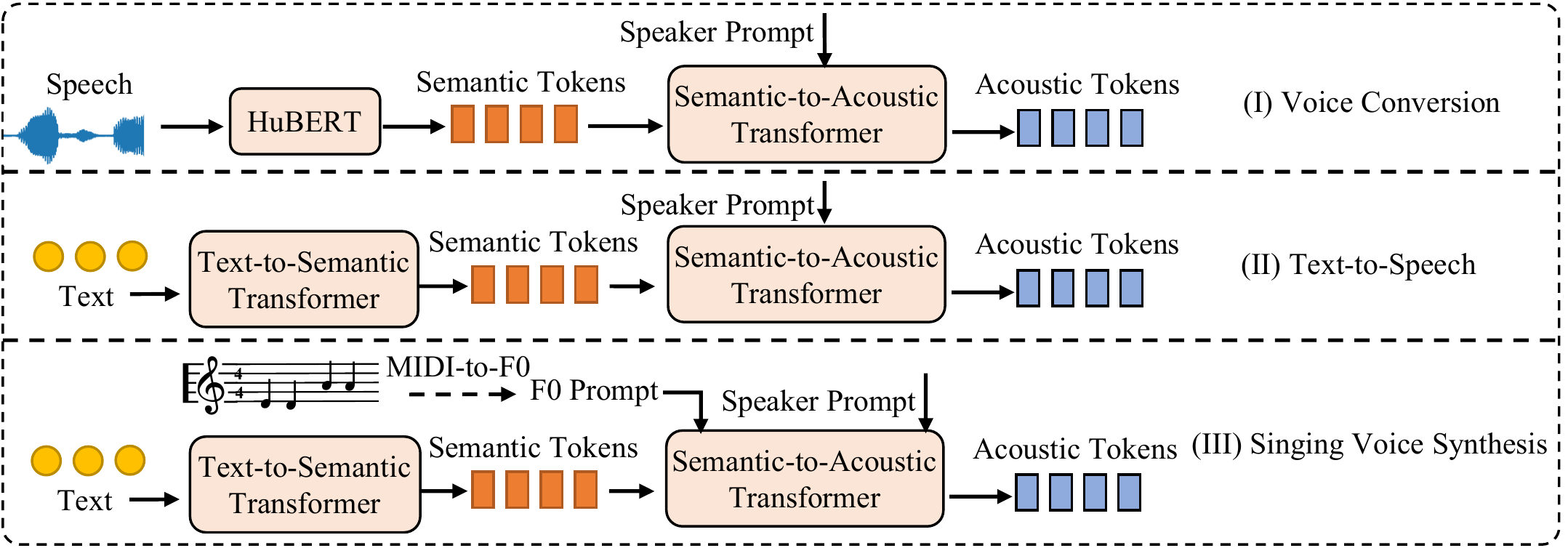}
    \caption{We introduce three exemplar applications, including voice conversion (VC), text-to-speech (TTS), and singing voice synthesis (SVS), that can be tackled by sharing a voice synthesis framework with semantic and acoustic tokens.}
    \label{fig:diagram}
  \end{figure*}

\subsubsection{Training}
During training text-to-semantic or semantic-to-acoustic transformers for stage $S_1, S_2$, we compute the cross-entropy (CE) loss with label smoothing between the generated and target units. Considering the synthesizer $S_3$, we train the enhanced vocoder with the weighted sum of the least-square adversarial loss, the feature matching loss, and the spectral regression loss on mel-spectrogram, where the training objective formulation and hyperparameters follow~\citet{kong2020hifi,lee2022bigvgan}. The major backbone of Make-A-Voice (i.e., both acoustic and generation stages) does not require any annotations, and thus training data could be scaled up to a large number of speakers, with various accents, diverse demographics, and heterogeneous recording conditions. It enables capturing acoustic diversity (speaker identity, emotion, prosody) in human voice, especially for zero-shot scenarios.

\subsubsection{Inference}
Make-A-Voice exhibits competitive advantages as a unified voice synthesis framework with a ``coarse-to-fine'' design. VC, TTS, and SVS can be tackled by re-synthesizing the semantic or acoustic representations with varying conditions:
\begin{itemize}[leftmargin=*]
    \item VC: speech sample is tokenized into semantic tokens, which are then transformed to acoustic tokens in target speaker given speaker prompt. As semantic and acoustic representations are produced with the same downsampling rate; thus, the input and converted speech share a common length. 
    \item TTS: phoneme sequences are translated to semantic tokens, which are converted into acoustic tokens given speaker prompt.
    \item SVS: phoneme sequences are translated to semantic tokens, which are transformed into acoustic tokens given the F0 prompt (from MIDI) and speaker prompt. Detailed information on the MIDI-to-F0 converter has been included in Appendix~\ref{midi}.
\end{itemize}

\section{Experiments}
\subsection{Experimental Setup}

\subsubsection{Data}
As illustrated in Table~\ref{data}, we separately train the backbone model on English and Chinese datasets since the semantic tokenizer (i.e., HuBERT) is mono-lingual, and we convert the sampling rate of all data to 16kHz. $S_1$ model is trained on text-semantic pairs, where we convert the text sequence into the phoneme sequence with an open-source grapheme-to-phoneme conversion tool~\citep{sun2019token}. $S_2$ and $S_3$ could be pre-trained on the audio-only self-supervised dataset to expand data distribution. During evaluation, we randomly choose 50 sentences to construct the zero-shot testing set for each application task, in which the voice used for prompting is never seen by the model at training, and it has to reproduce the characteristics from a single prompt example. We have attached more detailed information on the data configuration in Appendix~\ref{appendix:data}.

\begin{table}[ht]
    \centering
    \small
    \caption{Dataset usage in training and inference stages.}
    \scalebox{0.9}{
    \begin{tabular}{lcccc}
    \toprule
    Language & $S_1$             & $S_2/S_3$                     & Application           & Zero-Shot Testing \\
    \midrule
    English  & LibriTTS train & LibriLight                & TTS, VC               & LibriTTS test    \\
    Chinese  & OpenCPOP train & OpenCPOP+OpenSinger+CSMSC & SVS                   & M4Singer test    \\
    \bottomrule 
    \end{tabular}}
    \label{data}
    \end{table}

\subsubsection{Model Configurations}

For semantic representations, we apply different HuBERT models for the English and Chinese languages and use k-means to discretize 12th-layer embeddings into semantic tokens with a codebook of size 1000 and a total downsampling rate of 320. For acoustic representation, we train the SoundStream model with 12 quantization levels, each with a codebook of size 1024 and the same downsampling rate of 320. We take 3 quantization levels as the acoustic tokens, representing each frame as a flat sequence of tokens from the first, second, and third quantization layers. 

Autoregressive seq2seq text-to-semantic $S_1$ and semantic-to-acoustic $S_2$ models are both 12-layer transformers with an attention dimension of 1024 and the FFN dimension of 4096. As for unit-based vocoder $S_3$, we use the modified V1 version of BigVGAN. A comprehensive table of hyperparameters is available in Appendix~\ref{appendix:Model}.

\subsubsection{Training and Evaluation}
During training, we train $S_1$ and $S_2$ transformers respectively for 100K/500K steps using 4/32 NVIDIA V100 GPUs with a batch size of 2000 tokens for each GPU on the publicly-available \textit{fairseq} framework~\citep{ott2019fairseq}. Adam optimizer is used with $\beta_{1}=0.9, \beta_{2}=0.98, \epsilon=10^{-9}$. For the $S_2$ model, we crop the waveform to a random length of up to 8 seconds. $S_3$ model is optimized with a segment size of 8192 and a learning rate of $1 \times 10^{-4}$ until 500K steps using 4 NVIDIA V100 GPUs. During inference, we use beam search with beam size to $10$ of autoregressive decoding in both $S_1$ and $S_2$. 

We conduct a crowd-sourced human evaluation via Amazon Mechanical Turk, which is reported with 95\% confidence intervals (CI). We analyze in two aspects: style similarity (speaker, emotion, and prosody) and audio quality (clarity, high-frequency), respectively scoring SMOS and MOS. We further include objective evaluation metrics: Character Error Rate (CER) evaluates the intelligibility of the generated speech by transcribing it using a wav2vec ASR system. Speaker Cosine Similarity (Cos) and F0 Frame Error (FFE) measure the timbre and prosody similarity of synthesized and reference audio, respectively. More information has been attached in Appendix~\ref{appendix:Evaluation}.

\subsection{Text-to-Speech}

\begin{table}[ht]
    \centering
    \small
    \caption{Quality and style similarity of generated samples in zero-shot text-to-speech.}
    \begin{tabular}{lcccc}
    \toprule
    Model                       & MOS ($\uparrow$)   & SMOS ($\uparrow$)  & CER ($\downarrow$) & Cos ($\uparrow$) \\
    \midrule
    GT                          & 4.23$\pm$0.09   & /             &  0.030 &  / \\
    \midrule           
    Meta-StyleSpeech            &  3.84$\pm$0.08 & 3.76$\pm$0.09  &\textbf{0.065} & 0.73 \\
    \midrule
    \multicolumn{4}{l}{\textbf{Zero-Shot Text-to-Speech}} \\
    GenerSpeech                 &  3.99$\pm$0.08 & 3.77$\pm$0.08  &0.059 & 0.75 \\
    YourTTS                     &  3.89$\pm$0.08 & 3.72$\pm$0.06  &0.143 & 0.72 \\
    Make-A-Voice (TTS)          &  \textbf{4.04$\pm$0.07} & \textbf{3.81$\pm$0.08}  &0.068 & \textbf{0.77} \\
    \midrule 
    \multicolumn{5}{l}{Small-Scale Subjective Test} \\
    \midrule 
    VALL-E                &  3.92$\pm$0.12 & 3.81$\pm$0.07   &   /  &  /   \\
    SPEAR-TTS              &  3.98$\pm$0.06 & \textbf{3.84$\pm$0.06}  &   /  &  /   \\
    Make-A-Voice (TTS)     &  \textbf{4.05$\pm$0.10} & 3.83$\pm$0.08  &   /  &  /    \\
    \bottomrule 
    \end{tabular}
    \label{TTS}
    \end{table}



We compare the generated audio samples with other systems, including 1) GT, the ground-truth audio; 2) Meta-StyleSpeech~\citep{min2021meta}: the multi-speaker model fine-tuned with meta-learning; 3) GenerSpeech, a generalizable model with the global and local stylization for speaker and prosody style transfer; 4) YourTTS~\citep{casanova2022yourtts}: a zero-shot multi-speaker TTS model which is built upon VITS~\citep{kim2021conditional}; For easy comparison, the results are compiled and presented in Table~\ref{TTS}, and we have the following observations: 

1) For the intelligibility of the generated speech, Make-A-Voice (TTS) has achieved a CER of 0.068, comparable with other systems, indicating that Make-A-Voice (TTS) could generate accessible speech of good quality as previous non-autoregressive TTS families. 2) For audio quality, Make-A-Voice has achieved the highest MOS with scores of $4.04$ compared with the baseline models, demonstrating the effectiveness of the $S_3$ model in generating high-fidelity waveforms. 3) Regarding style similarity, Make-A-Voice scores the SMOS of $3.81$. The objective results of cosine similarity further show that Make-A-Voice (TTS) surpasses the state-of-the-art models in transferring the style of custom voices. Informally, Make-A-Voice is optimized in a large amount of self-supervised data, which contains many speakers with various accents, diverse demographics, and heterogeneous recording conditions, to improve robustness and generalization in zero-shot scenarios. We further include the discussion on generation speed in Appendix~\ref{limit}.

Using the examples provided on its demo page, we also compare Make-A-Voice with SPEAR-TTS and VALL-E in a small-scale subjective test. We synthesize 24 utterances using the same transcripts and prompts and conduct a subjective test with the same protocol described above. Table~\ref{TTS} shows that, in these examples, Make-A-Voice achieves considerably higher naturalness (MOS 4.05) than VALL-E (3.92) and SPEAR-TTS (3.98) and comparable style similarity in zero-shot synthesis.

\subsection{Voice Conversion}

\begin{table}[ht]
    \centering
    \small
    \caption{Quality and style similarity of generated samples in zero-shot voice conversion.}
    \begin{tabular}{lcccc}
    \toprule
    Model                   & MOS ($\uparrow$) & SMOS ($\uparrow$)    & Cos ($\uparrow$) \\
    \midrule
    GT                     & 4.26$\pm$0.06  & /             &   /  \\
    \midrule           
    NANSY                 &  3.89$\pm$0.08  & 3.73$\pm$0.10  & 0.68 \\
    PPG-VC                &  3.97$\pm$0.06  & \textbf{3.80$\pm$0.07}  & 0.78 \\
    \midrule
    \multicolumn{4}{l}{\textbf{Zero-Shot Voice Conversion}} \\
    Make-A-Voice (VC)     &  \textbf{4.07$\pm$0.06}  & 3.77$\pm$0.07    & \textbf{0.80} \\
    \bottomrule 
    \end{tabular}
    \label{vc}
    \end{table}

In this part, we describe Make-A-Voice (VC), where we perform zero-shot voice conversion by tokenizing speech using the HuBERT model $S_1$, and leverage new voice for prompting in semantic-to-acoustic translation $S_2$. We compare the generated audio samples with other systems, including 1) GT, the ground-truth audio; 2) NANSY~\citep{choi2022nansy++}: the unified framework of synthesizing and manipulating voice signals from analysis features; 3) PPG-VC~\citep{liu2021any}: the voice conversion model based on phonetic posterior-gram. The results are presented in Table~\ref{vc}, and we have the following observations: 

1) Make-A-Voice (VC) scores the comparable overall SMOS of $3.77$ with baseline, showing that it excels at converting speaker identity even in a zero-shot scenario, attributing to the scalable training data covering diverse speakers with various accents. 2) For audio quality, it presents high perceptual quality with outperformed MOS evaluation. To conclude, Make-A-Voice (VC) converts the timbre with better naturalness and comparable speaker similarity compared to baseline models, even though the model is trained without any text transcript paired with audio recordings. 


\subsection{Singing Voice Synthesis}

\begin{table}[ht]
    \centering
    \small
    \caption{Quality and style similarity of generated samples in zero-shot singing voice synthesis.}

    \begin{tabular}{lcccc}
    \toprule
    Model                   & MOS ($\uparrow$)     & SMOS ($\uparrow$)      & Cos ($\uparrow$) & FFE ($\downarrow$) \\
    \midrule
    GT                     & 4.08$\pm$0.08  &  /   &   /  &  /   \\
    \midrule           
    FFT-Singer             &  3.83$\pm$0.09 & 3.91$\pm$0.08    & 0.93    & 0.17 \\
    Diffsinger             &  3.98$\pm$0.07 & \textbf{3.98$\pm$0.07}  & \textbf{0.94}    & 0.08\\
    \midrule
    \multicolumn{4}{l}{\textbf{Zero-Shot Singing Voice Synthesis}} \\
    Make-A-Voice (SVS)     &  \textbf{3.99$\pm$0.08} & 3.96$\pm$0.07  & 0.88    &  \textbf{0.05}   \\
    \bottomrule 
    \end{tabular}
    \label{svs}
    \end{table}

Make-A-Voice (SVS) conducts zero-shot singing voice synthesis by leveraging the $S_1$ model for text-to-semantic conversion, together with the F0-guided acoustic model $S_2$ and generator $S_3$. The limited amount of labeled data~\citep{choi2022nansy++} and unpleasant glitches~\citep{huang2022singgan} are two major challenges for singing voice synthesis, which we address as follows: 1) For data scarcity, Make-A-Voice leverages a large amount of unlabeled speech and singing voice data for modeling the acoustic diversity (speaker, prosody, and style) in voice, which only requires a small amount of parallel data to learn the text-semantic mapping. 2) To reduce glitches, we include the F0-driven source excitation to stabilize long-continuous waveform modeling in the $S_3$ stage.

We compare the generated singing voice samples with other systems, including 1) GT, the ground-truth audio, which is the upper limit for SVS; 2) Diffsinger~\citep{liu2022diffsinger}, an acoustic model based on the diffusion probabilistic model, 3) FFT-Singer, which generates mel-spectrograms through stacked feed-forward transformer blocks. As illustrated in Table~\ref{svs}, and we have the following observations: 

1) In terms of prosody similarity, Make-A-Voice (SVS) outperforms the baseline system by a large margin, showing distinct superiority in terms of FFE objective evaluation. Make-A-Voice (SVS) can resemble the prosodic style of the F0 prompt and demonstrates its precise pitch reconstruction. 2) Informally, a gap from baseline models regarding singer cosine similarity could be witnessed, since they share a static number of singers during training and inference. In contrast, the voice used for prompting is never seen by Make-A-Voice (SVS) at training, which is more challenging to reproduce the characteristics from a single prompt in a zero-shot manner, especially for singing voice.

\subsection{Analysis and Ablation Studies}

To verify the effectiveness of several designs in Make-A-Voice, including the usage of the HuBERT layer for semantic tokens, and the voice reconstruction from acoustic tokens, we conduct ablation studies and discuss the key findings as follows.

\subsubsection{From which layer should the semantic features be extracted?} 

\begin{wraptable}{r}{7cm}
    \centering
    \small
    \vspace{-5mm}
    \caption{Ablation studies.}
    \vspace{2mm}
    \begin{tabular}{lccc}
    \toprule
    Model            & CER ($\downarrow$)  & STOI ($\uparrow$) & MCD ($\downarrow$) \\
    \midrule
    HuBERT-10        & 0.54 & /    & /   \\
    HuBERT-11        & 0.44 & /    & /   \\
    HuBERT-12        & \textbf{0.39} & /    & /   \\
    \midrule
    $S_3$: SoundStream  & /   & 0.92   & 1.90 \\
    $S_3$: Unit Vocoder & /   & \textbf{0.93}   & \textbf{1.56} \\
    \bottomrule 
    \end{tabular}
    \label{ablation}
    \end{wraptable}
Recently, ~\citet{shah2021all,choi2021neural} show that it is the output from the middle layer of the SSL model (e.g., Wav2vec 2.0~\citep{baevski2020wav2vec}, HuBERT~\citep{hsu2021hubert}) that has the most relevant characteristics to pronunciation. Thus, we train semantic unit-based vocoders~\citep{polyak2021speech} with representations clustered from (the 10th, 11th, and 12th) of the 24 HuBERT layers and report the CER of synthesized speech. In light of empirical observation in Table~\ref{ablation}, we find that the 12th layer feature of HuBERT enjoys rich linguistic content information with the highest intelligibility, and thus we discretize 12th layer embeddings into semantic tokens.

\subsubsection{Why use unit-based vocoder instead of original SoundStream decoder?}  \label{ablation:vocoder}

As illustrated in Table~\ref{ablation}, replacing the unit-based vocoder with a SoundStream decoder for voice synthesis has witnessed a distinct degradation of perceptual quality, proving that the codebook mismatch for the SoundStream decoder between training (12 quantization levels) and inference (3 levels) hurts reconstruction performance. In contrast, a neural vocoder could refine the coarse-grained acoustic tokens and generate waveforms with increasing details.

\subsubsection{Zero-shot transfer beyond speaker identity} 

This section presents how our approach could be extended beyond, including cross-lingual timbre transferring, generating coherent emotion, and noise continuations. We have attached the information on testing data in Appendix~\ref{appendix:data}. As shown in the demo page, we find that 1) Make-A-Voice can preserve the \textbf{emotion} in the prompt at a zero-shot setting, even if the model is not fine-tuned on an emotional TTS dataset; 2) Make-A-Voice effectively reproduces the characteristics from a \textbf{cross-lingual} style prompt, which has not been seen during training; and 3) In a noisy environment, the model also presents the acoustic consistency and maintain the \textbf{noise conditions} in the prompt.
\section{Conclusion}

In this work, we proposed Make-A-Voice, a unified framework for synthesizing and manipulating voice signals from discrete representations. Make-A-Voice enjoyed the ``coarse-to-fine'' design to model the human voice, which involved three stages: 1) semantic: from text/speech to determine semantic meanings, 2) acoustic: from semantic tokens to introduce acoustic conditions, 3) generation: from acoustic tokens to high-fidelity waveforms. Various applications for voice synthesis could be tackled by sharing a unified framework with semantic and acoustic tokens, and we introduced text-to-speech (TTS), voice conversion (VC), and singing voice synthesis (SVS). Experimental results demonstrated that Make-A-Voice offered notable benefits as a unified voice synthesis framework: 1) Data scalability: the backbone (i.e., both acoustic and generation stage) did not require any annotations, and thus training could be scaled up regarding data usage. 2) Controllability and conditioning flexibility: various conditioning mechanisms were investigated by re-synthesizing the semantic or acoustic representations with varying control signals (e.g., speaker/F0 prompt). For future work, we will verify the effectiveness in more general scenarios such as multilingual generalization. The discussions on limitations and potential negative impacts are included in Appendix. We envisage that our work will serve as a basis for future voice synthesis studies.
\clearpage
\bibliographystyle{neurips_2023}
\bibliography{neurips_2023}

\newpage
\appendix

\section{Data} \label{appendix:data}
In this section, we describe details of the data usage in training and evaluating Make-A-Voice.

\begin{itemize}[leftmargin=*]
    \item English: We use LibriLight~\citep{kahn2020libri} as the training data which contains 60K hours and around 7k speakers of unlabelled speech from audiobooks in English. During the evaluation, we utilize the test split in LibriTTS~\cite{panayotov2015librispeech} as a zero-shot testing dataset, where the voice used for prompting is never seen at training.
    \item Chinese: We use 5-hour OpenCPOP~\citep{wang2022opencpop} with one female singer, and 50-hour multi-singer dataset OpenSinger~\citep{huang2021multi} as the singing voice training data. In addition, Chinese standard Mandarin speech corpus~\footnote{\url{https://www.data-baker.com/open_source.html}} is also included during training. During the evaluation, we utilize the large human-labeled M4Singer~\citep{zhang2022m4singer} with a benchmark in downstream singing voice synthesis.
    \item Beyond speaker identity transfer: We prepare the zero-shot testing data from ESD dataset~\citep{zhou2021seen}, and prepare noise audio clips in the categories of ``speech'' sampled from MUSAN dataset~\citep{snyder2015musan}.
\end{itemize}

\section{Model Configurations} \label{appendix:Model}

We list the model hyper-parameters of Make-A-Voice in Table~\ref{tab:hyperparameters}.

\begin{table}[h]
    \small
    \centering
    \begin{tabular}{l|c|c}
    \toprule
    \multicolumn{2}{c|}{Hyperparameter}   & Make-A-Voice \\ 
    \midrule
    \multirow{6}{*}{Stage $S_1$: Text-to-Semantic Transformer} 
    &Transformer Layer         &    12     \\
    &Transformer Embed Dim                   &   1024  \\    
    &Transformer Attention Headers           &  16   \\  
    &Transformer FFN Embed Dim              &  4096   \\    
    &Decoder Dictionary Length          &   1000 \\
    &Number of Parameters               & 321.3M \\
    \midrule
    \multirow{6}{*}{Stage $S_2$: Semantic-to-Acoustic Transformer} 
    &Transformer Layer                      &    12     \\
    &Transformer Embed Dim                   &   1024  \\    
    &Transformer Attention Headers           &  16   \\  
    &Transformer FFN Embed Dim              &  4096   \\    
    &Decoder Dictionary Length              &   1024 \\ 
    &Number of Parameters               & 323.3M \\
    \midrule
    \multirow{4}{*}{Stage $S_3$: Unit-based vocoder} 
    &Upsample Rates                     &    [5, 4, 2, 2, 2, 2]     \\
    &Hop Size                 &   320  \\    
    &Upsample Kernel Sizes          &  [9, 8, 4, 4, 4, 4]   \\
    &Number of Parameters               & 121.6M \\
    \midrule 
    \multicolumn{2}{c|}{Total Number of Parameters}   &  776.2M \\
    \bottomrule
    \end{tabular}
    \vspace{2mm}
    \caption{Hyperparameters of Make-A-Voice.}
    \label{tab:hyperparameters}
    \end{table}

\section{Applications} \label{appendix:application}

\subsection{MIDI-to-F0 Converter} \label{midi}
Singing voice synthesis (SVS) is a task that generates singing voices from the given music score and lyrics like human singers. Following~\citep{liu2022diffsinger,zhang2022m4singer}, the SVS system typically includes the MIDI-to-F0 converter to predict F0 explicitly. Though the SVS system can be further improved with the direct MIDI condition and implicit F0 prediction, this is beyond our focus.

\subsection{Unit-based Vocoder} \label{vocoder}

\begin{figure*}[ht]
    \centering
    \vspace{-2mm}
    \includegraphics[width=.4\textwidth]{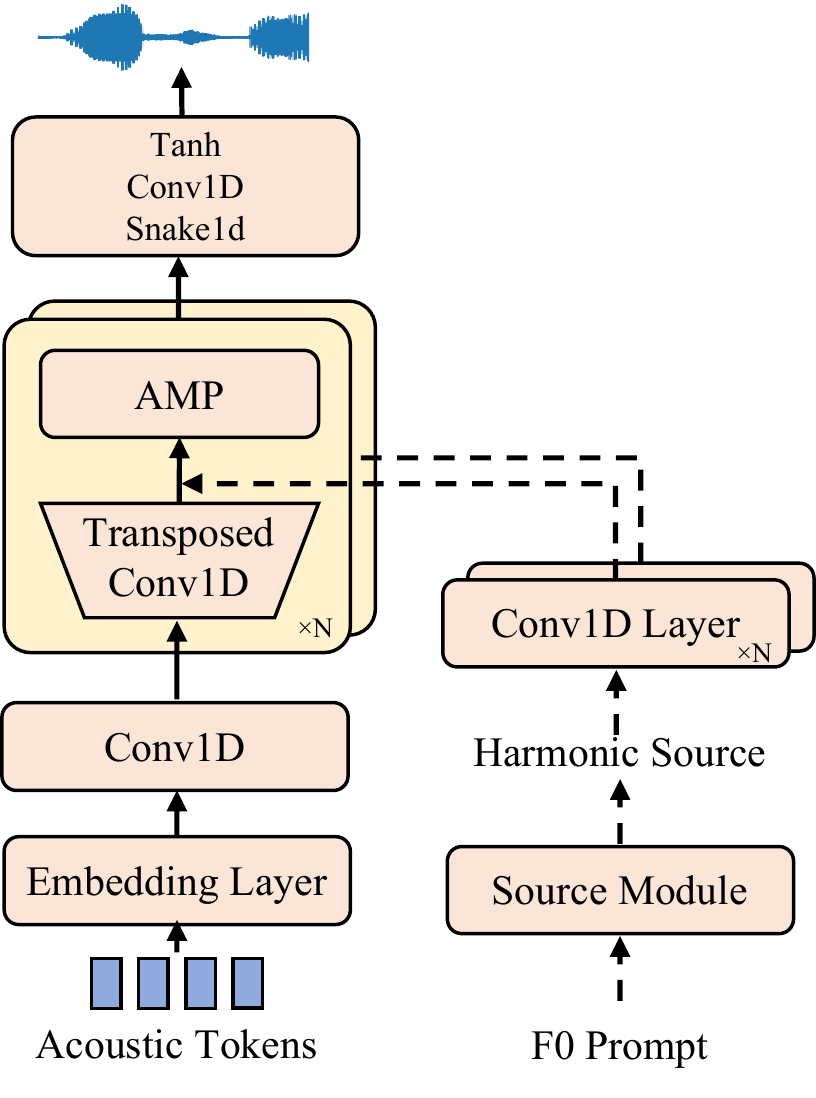}
    \vspace{-2mm}
    \caption{Overview of the unit-based vocoder. The F0 auxiliary input denoted with dotted lines is included only in singing voice synthesis.}
     \vspace{-2mm}
    \label{fig:arch}
  \end{figure*}
  
The generator of the unit-based vocoder is built from a set of look-up tables (LUT) that embed the discrete representation, and a series of blocks composed of transposed convolution and a residual block with dilated layers.

    \section{Evaluation}\label{appendix:Evaluation}

    \subsection{Subjective Evaluation}  \label{Subjective}

    For audio quality evaluation, we conduct the MOS (mean opinion score) tests and explicitly instruct the raters to ``\textit{(focus on examining the audio quality and naturalness, and ignore the differences of style (timbre, emotion, and prosody).)}". The testers present and rate the samples, and each tester is asked to evaluate the subjective naturalness on a 1-5 Likert scale.
    
    For style similarity evaluation, we explicitly instruct the raters to ``\textit{(focus on the similarity of the style (timbre, emotion, and prosody) to the reference, and ignore the differences of content, grammar, or audio quality.)}". In the SMOS (similarity mean opinion score) tests, we paired each synthesized utterance with a ground truth utterance to evaluate how well the synthesized speech matches that of the target speaker. Each pair is rated by one rater. 

\begin{figure}[!h]
	\centering
    \subfigure[Screenshot of MOS testing.]
    {
    \includegraphics[width=0.9\textwidth]{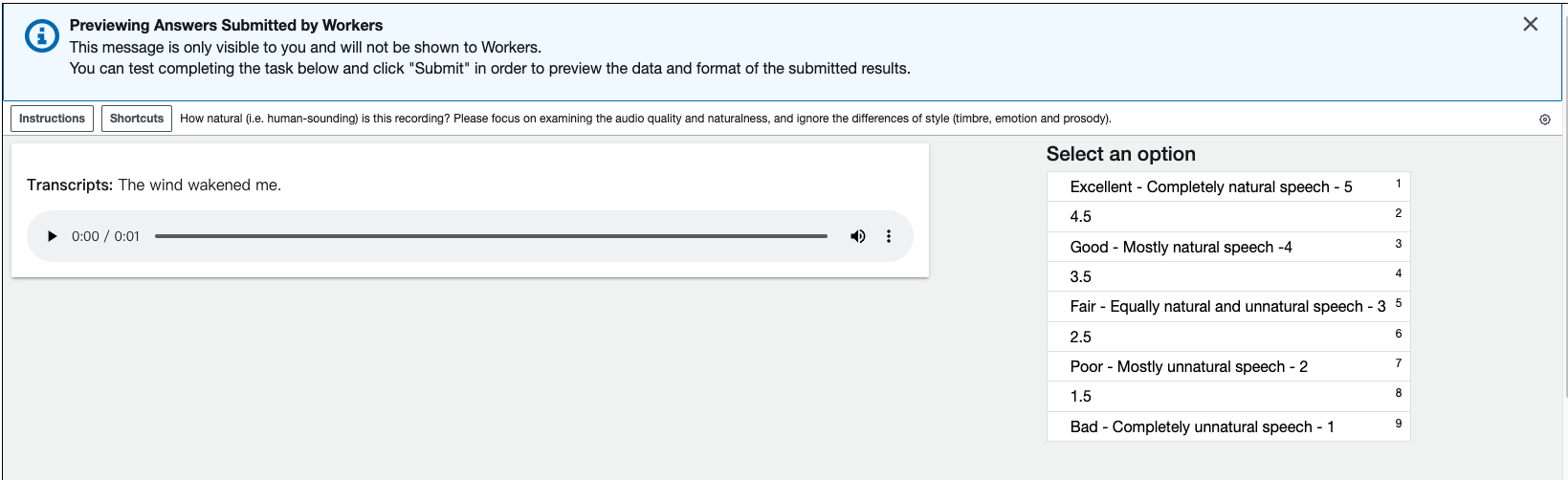}
    }
    \subfigure[Screenshot of SMOS testing.]
    {
    \includegraphics[width=0.9\textwidth]{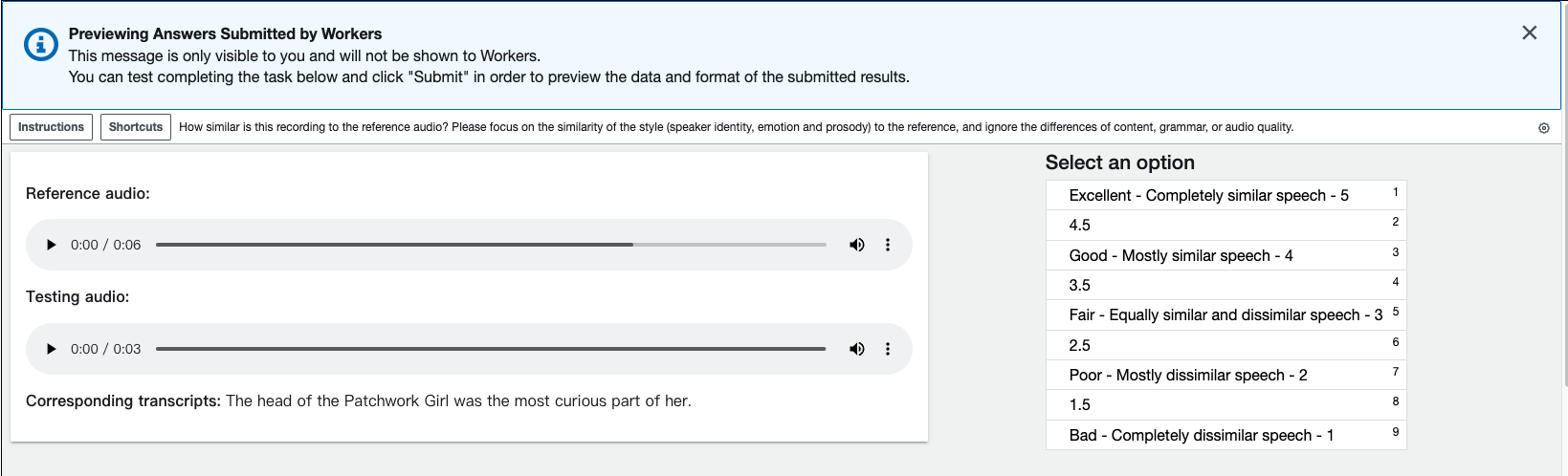}
    }
	\caption{Screenshots of subjective evaluations.}
	\label{fig:screenshot_eval}
\end{figure}

    Our subjective evaluation tests are crowd-sourced and conducted by 20 native speakers via Amazon Mechanical Turk. The screenshots of instructions for testers have been shown in Figure~\ref{fig:screenshot_eval}. We paid \$8 to participants hourly and totally spent about \$600 on participant compensation. A small subset of speech samples used in the test is available at \url{https://Make-A-Voice.github.io/}.

\subsection{Objective Evaluation}  \label{Objective}

Cosine similarity is an objective metric that measures speaker similarity among multi-speaker audio. We compute the average cosine similarity between embeddings extracted from the synthesized and ground truth embeddings to measure the speaker similarity performance objectively. 

Character Error Rate (CER) evaluates the faithfulness to the input transcript by transcribing the synthesized utterances using a wav2vec ASR system. 

F0 Frame Error (FFE) combines voicing decision error and F0 error metrics to capture F0 information. 

Mel-cepstral distortion (MCD) measures the spectral distance between the synthesized and reference mel-spectrum features.

Short-time objective intelligibility (STOI) assesses the denoising quality for speech enhancement.







\end{document}